# Role of Domain Walls on Imprint and Fatigue in HfO$_2$-Based Ferroelectrics

Muting Xie [a,b,‡], Hongyu Yu [a,b,‡], Binhua Zhang [a,b], Changsong Xu[a,b]*, and Hongjun Xiang[a,b]†

[a]Key Laboratory of Computational Physical Sciences (Ministry of Education), Institute of Computational Physical Sciences, State Key Laboratory of Surface Physics, and Department of Physics, Fudan University, Shanghai 200433, China

[b]Shanghai Qi Zhi Institute, Shanghai 200030, China

*E-mail: csxu@fudan.edu.cn

†E-mail: hxiang@fudan.edu.cn

‡These authors contributed equally to this work.

Abstract

HfO$_2$-based ferroelectric materials are promising for the next generation of memory devices, attracting significant attention. However, their potential applications are significantly limited by fatigue and imprint phenomena, which affect device lifetime and memory capabilities. Here, to accurately describe the dynamics and field effects of HfO$_2$, we adopt our newly developed DREAM-Allegro network scheme and develop a comprehensive machine-learning model for HfO$_2$. Such model can not only predict the interatomic potential, but also predict Born effective charges. Applying such model, we explore the role of domain dynamics in HfO$_2$ and find that the fatigue and imprint phenomena are closely related to the so-called E-path and T-path switching pathways. Based on the different atomic motions in the two paths, we propose that an inclined electric field can sufficiently suppress fatigue and enhancing the performance of HfO$_2$-based ferroelectric devices.

Hafnium dioxide (HfO$_2$)-based materials, particularly Hf$_x$Zr$_{1-x}$O$_2$, are considered promising candidates for advanced complementary metal–oxide–semiconductor (CMOS) technology [1–3]. These ferroelectric materials shows good compatibility with CMOS processes and may retain their ferroelectric properties even at the nanoscale [4–8]. Such advantages make the HfO$_2$-based materials promising in next-generation information technology and received great attention [9–16]. However, despite the promising applications, the applications of HfO$_2$-based ferroelectrics still suffer from technical problems, such as fatigue and imprint. Fatigue in ferroelectric materials refers to the phenomenon where the polarization decreases with repeated cycling of an electric field (Fig. 1(b)), which limits the operational lifespan of the ferroelectric devices [17,18]. Whereas the imprint effect is characterized by a voltage shift in the polarization hysteresis loop, causing an increase in operating voltage (Fig. 1(a)). Imprints can thus lead to an increase in writing failures in ferroelectric memory devices [19,20]. Apparently, fatigue and imprints can significantly affect the performance and reliability of ferroelectric devices.

Although fatigue and imprint are common in ferroelectrics, their origins are not clearly understood. Point defects, especially charged oxygen vacancies, are considered the most likely cause for both imprint and fatigue in different researches in HfO$_2$-based ferroelectrics [10,12,21–26]. Other defects, such as structural phase transitions, also have experimental supports [18,27]. On another aspect, domain walls (DWs) typically play an important role in ferroelectrics due to the widely accepted model of nucleation and expansion [28–30]. The relatively low symmetry of Pca2$_1$ phase [31–33] of HfO$_2$ results in multiple types of DWs [34,35] and switching pathways [9,36], indicating complex influences of domains and their dynamics in HfO$_2$-based ferroelectrics. Although certain static properties of domains in HfO$_2$ have been studied [37–39], the dynamics of domains and their relationship with imprint and fatigue have received little attention.

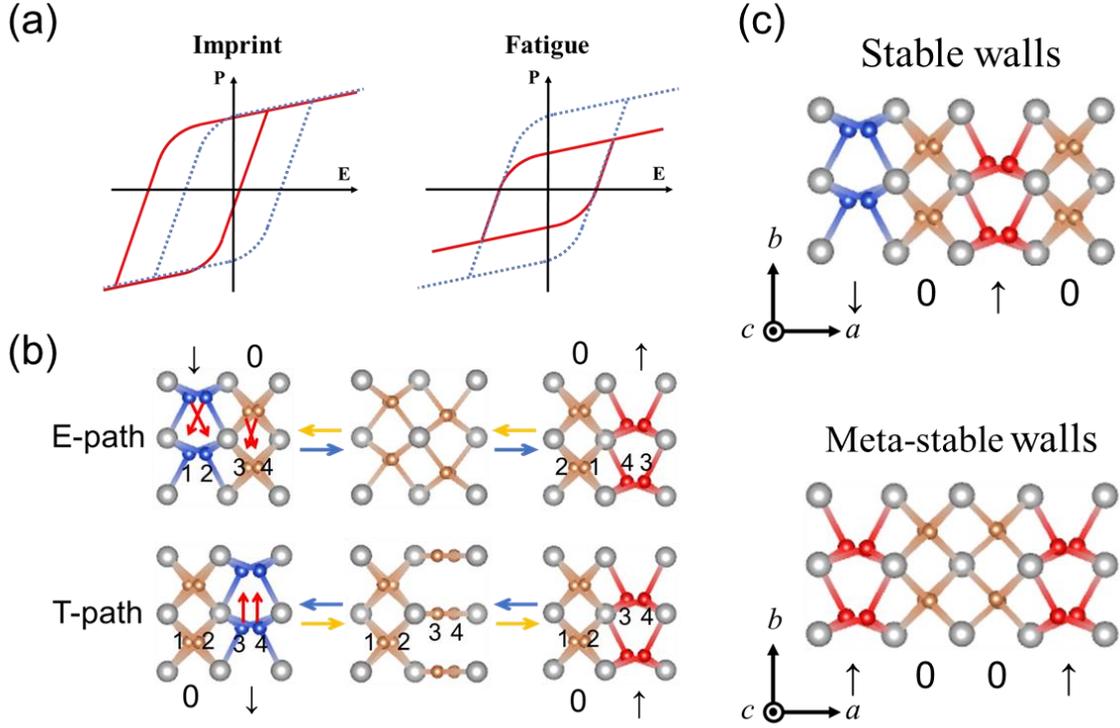

**Figure 1** (a) Schematics of imprint and fatigue phenomena. (b) Schematics of E and T switching pathways for $HfO_2$. Grey balls for hafnium. Grown, red, blue balls for different oxygens. The blue (yellow) arrows represent the transition trend of application of an electric field along positive (negative) $b$ direction. (c) Two typical DWs that are coined stable walls and meta-stable walls. The stable DWs resemble the Pbca phase, while the meta-stable DWs are characterized by the adjacency of two '0's.

In this Letter, we address the issues of imprint and fatigue in $HfO_2$-based systems by focusing on domain structures and domain dynamics. We develop a new neural network called DREAM [40], i.e., a generalized dielectric response equivariant atomistic multitask framework, based on Allegro [41]. The DREAM network can compute both interatomic potentials and the Born effective charges (BECs). Applying DREAM, we develop machine learning model for $HfO_2$-based systems. Details of the model are shown in part A of the SM(supplementary materials) [42]. By performing molecular dynamics simulations, we reveal the critical role of DWs in imprint and fatigue phenomena. Specifically, the dynamics of domains are represented by two main switching pathways: the E-path, which can induce ferroelectricity and imprint depending on domain structures, and the T-path, which generates DWs and causes fatigue and imprint effects. Moreover, we propose a strategy of applying inclined electric field to effectively suppress fatigue in $HfO_2$-based materials.

We start from the basic domain structures in $HfO_2$-based systems. A unit cell of ferroelectric $HfO_2$, belonging to the space group $Pca2_1$, comprises 4 hafnium atoms and 8 oxygen atoms and can be divided into a polar half-cell and a nonpolar half-cell. The orientation of polarizations is along the '*b*' axis and we view structures along the '*c*' axis (Fig. 1). We use '0' to signify the nonpolar half-cell, in which the oxygen atoms are located between two adjacent Hf layers. The polar half-cell are denoted by marks of '↑' and '↓', where the former is for the oxygen atoms located on the lower side and '↓' for those on upper side. With this notation, domain structures can be represented with '↑', '0' and '↓'. Uniform domains can be represented by regions with periodic (0↑)s or (0↓)s. Two types of DWs are found important: the stable DW (↓0↑) and the meta-stable DW (↑00↑), as illustrated in Fig. 1(c). Stable DW (↓0↑) exhibit energy of 2 meV/Å$^2$ lower than the uniform domains, whereas meta-stable DW (↑00↑) is 20 to 40 meV/Å$^2$ higher, depending on the applied electric field. Transiently appearing unstable structures, such as (↑00↓) and (↑↓), will not be discussed as they can only appear in non-equilibrium states. See details of the kinds of DWs in part F of the SM [42].

**Dynamics of domains.** It is found, during our simulations, that the dynamics of domains can be summarized into two pivotal ferroelectric switching pathways, i.e., E-path and T-path. The E-path process involves a switching between polar and nonpolar half-cells, with all oxygen atoms moving in the direction opposite to the electric field. Conversely, the T-path process only affects the polarized oxygen atoms, which move through the Hf plane, leaving the nonpolar oxygen atoms unchanged. The switching process of E-path and T-path are shown in Fig. 1(b). The notation used for the T-path is consistent with the notation in Evgeny's work [9]. The simulations show that the critical electric field for E-path initiation ($E_E$) is about 15 MV/cm for $HfO_2$ at T = 1 K, which is consistent with previous DFT calculations [43]. The critical electric field for T-path initiation ($E_T$) is approximately 22 MV/cm for $HfO_2$. Previous studies [9,11,36,44,45] have explored both the E-path and T-path to certain degrees.

Note that due to the periodic boundary conditions, only relative polarization can be calculated, which requires assessment based on the specific switching pathways. In this study, the up and down polarizations are defined based on the E-path. That is to say, if an electric field is applied in the upward direction, the T-path switches from the up (↑) to the down (↓) polarization.

Switching pathways, together with domain structures, can accurately represent domain dynamics. According to our MD simulations, the macroscopic behavior of the E-path, which reverses the polarization, resembles a wavefront formed by the displacement of oxygen atoms. This wavefront propagates freely across the domain until it encounters the DWs and is stopped. When a row of oxygen atoms undergone E-path switching under an electric field without DWs, it quickly induced adjacent E-paths, due to the instability of the boundary. Such behavior can be illustrated by the sequence of (0↓0↓0↓0↓) → (↑00↓0↓0↓) → (↑0↑00↓0↓) → (↑0↑0↑00↓) → (↑0↑0↑0↑0). While in the presence of DWs, the E-path sequence is as follows: (↑0↑0↑0↑0↓) ↔ (0↓0↓0↓00↓), oscillating between the stable DWs (↓0↑) and the meta-stable walls (↓00↓). As E-path can only spread inside a specific domain, it cannot create or annihilate DWs.

On the other hand, T-paths can create or annihilate DWs, acting as a defect generator. T-paths can occur within a domain or at a DW, requiring only a single row of polar half-cells. When a T-path switching event occurred within a domain, it would create a new DW, demonstrated by the sequence of (0↓0↓0↓0) → (0↓0↑0↓0). Although T-path is also capable of annihilating DWs, there are factors, which will be detailed in the fatigue part, that favors the creation process. Unlike the E-path, the T-path cannot induce adjacent T-paths. In the context of a T-path, two nearby rows of polar half-cells are separated by a row of nonpolar half-cells, thus resulting in a subtle correlation between adjacent T-paths. It's noteworthy that the E-path is the only genuine ferroelectric switching pathway capable of producing a ferroelectric hysteresis loop. For electric fields less than $E_T$, the E-path dominates. Once the electric field surpasses $E_T$, T-paths are followed by E-paths, leading to oxygen atom displacement [11], which is permissible only in PBC.

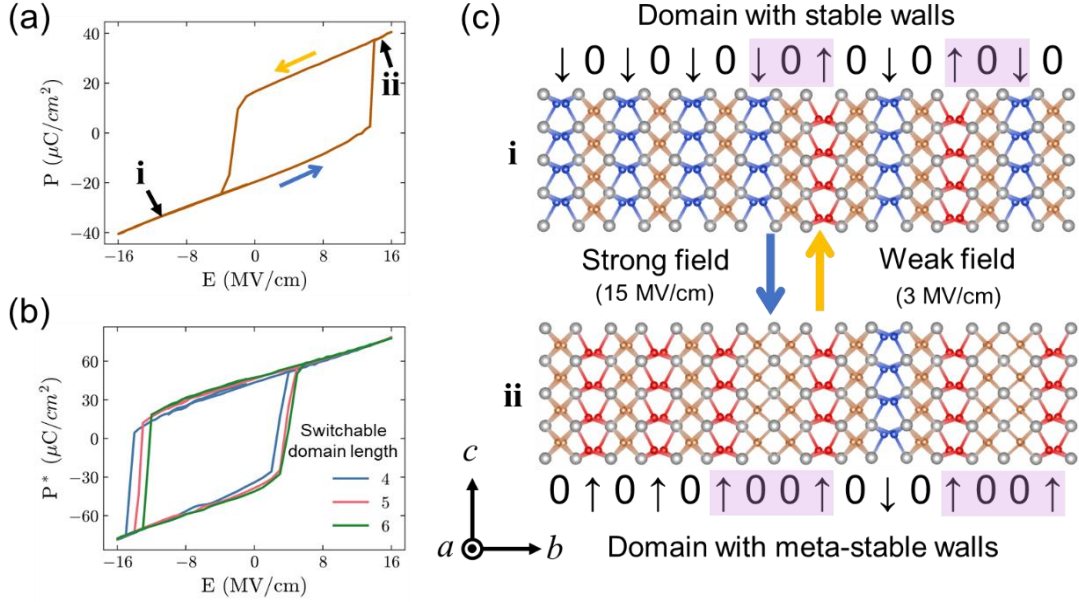

**Figure 2**. (a) Imprint-like biased hysteresis loop of $HfO_2$. (b) The biased hysteresis loops for different lengths of switchable domains, where the polarization of the switchable domain is considered. (c) Illustration of stage i and stage ii in the process of panel (a).

**Imprint phenomenon and its origin.** Imprint-like hysteresis loop, as shown in Fig. 2(a)(b), can be observed in systems with DWs. In MD simulations, we construct supercells containing a polar half-cell of $Pca2_1$ phase and a nonpolar half-cell of $Pbca$ phase to maintain DWs. Initially, we set the polarization of the $Pca2_1$ half-cell pointing downward and simulate the P-E loop. As shown in Fig. 2(a), remarkable reduction from 15 MV/cm to 3 MV/cm in the critical field is observed for polarization being switched from up to down. Conversely, the critical field for switching from down to up remained 15 MV/cm. Furthermore, it is found that longer domain length, i.e., lower density of DWs, of the $Pca2_1$ half-cell results in the formation of even more severely biased hysteresis loops, as shown in Fig. 2(b). Figure 2(b) also indicates that, in systems being initially up polarized, a reduction in domain length led to an increase in the critical fields for switching from up to down and a decrease in the critical fields for switching back to up.

The imprint-like hysteresis loops can be explained by the dynamics of DWs, as shown in Fig. 2(c). Initially, in stage i, where the polarized domain is constrained by stable DWs, applying an electric field transforms the system to stage ii, where the polarized domain is constrained by meta-stable DWs. As meta-stable walls are less stable than uniform domains, a strong electric field of approximately 15 MV/cm is required to induce E-path and switch from stage i to stage ii, while a weaker field of 3 MV/cm suffices to switch from stage ii to stage i, resulting in a biased hysteresis loop resembling an imprint. The impact of DWs depends on the distance between two proximate DWs; as the domain length between DWs diminishes, the DW density increases, and the distinction between stable and meta-stable DWs becomes more pronounced, giving rise to the observed tendencies in Fig. 2(b).

The imprint effect can be further explained by the thermally induced T-path and the movement of DWs. Although the energy barrier for the T-path is higher than that for the E-path, thermally induced T-paths may occur at high temperatures over a long duration. If a uniformly polarized down domain is baked for a sufficient period, it divides into several discontinuous polarized down domains separated by stable DWs. The longer the duration, the more T-paths occur, shortening the average length of the continuous domain and resulting in more significant imprint effects. The recovery of the imprint can be explained by the movement of DWs, as discussed in parts C and D of the SM [42].

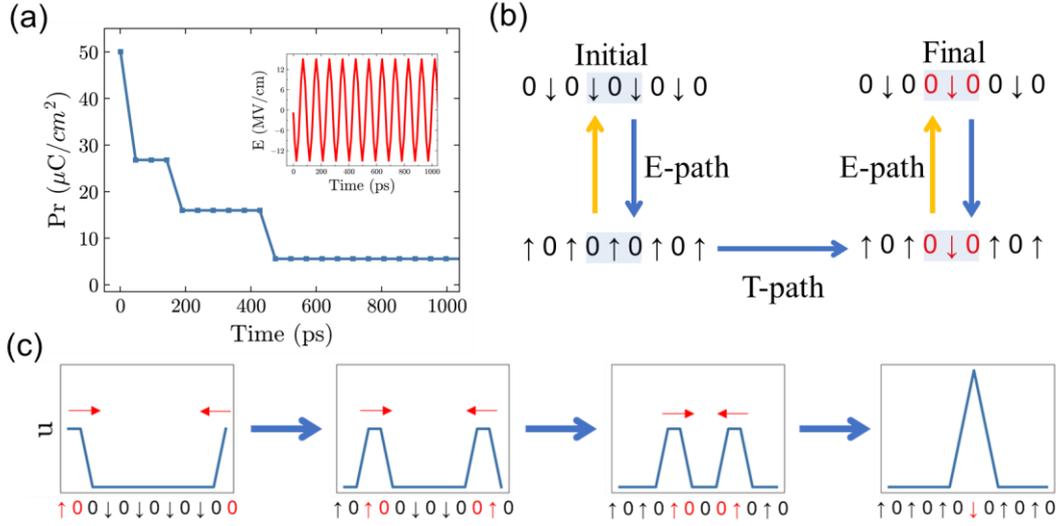

**Figure 3.** (a) The fatigue phenomenon observed in simulations at T = 20 K, showing the decrease in switchable polarization of $HfO_2$ over time. (b) Schematics illustrating the origin of fatigue from a domain perspective. (c) Schematics depicting E-path-induced T-paths, where 'u' represents the displacement of oxygen atoms with respect to the previous structure.

**Fatigue phenomenon and its origin.** During the simulation, a notable reduction in polarization was observed, implying the emergence of fatigue. The simulation begins with a single-domain 20×4×4 supercell containing 3840 atoms and ends in a disordered domain with DWs, with polarization decreasing from 50 to around 5 μC/cm². The simulations are conducted at T = 20 K, with switchable polarizations measured using additional NVT simulations to avoid temperature influence. As shown in Fig. 3(a), the polarization drops to 50% of its initial value immediately after the start of the simulation and stabilizes at 10% after 450 ps. At higher temperatures, the system loses its spontaneous polarization even faster, consistent with experimental observations that pure $HfO_2$ does not exhibit ferroelectricity at room temperature [46]. This simulation ends in a disordered domain containing DWs, aligning well with Cheng's experiment [27].

The origin of the observed fatigue can be traced to E-path-induced T-paths, unlike the thermo-induced T-paths seen in imprint processes. E-paths create wavefronts formed by the displacement of oxygen atoms. These wavefronts propagate freely within a domain but stop at DWs. When two E-path wavefronts meet, the resultant oxygen atom displacement becomes excessively large, leading to the formation of T-path and the occurrence of fatigue, as illustrated in Fig. 3(c). Longer domains and higher temperatures increase the probability of E-path wavefronts encountering each other before being stopped by DWs, leading to shorter average domain lengths and weaker polarization at the end of the fatigue process.

Although both imprint and fatigue involve T-paths, their mechanisms differ, leading to different phenomena. E-path-induced T-paths are more efficient in forming DWs than thermo-induced T-paths, leading to a more sufficient decrease of polarization in fatigue than that in imprint. In imprint processes, T-paths are constrained to a specific direction (e.g., from up to down for domains initially polarized upwards), resulting in biased hysteresis loops. In contrast, fatigue, caused by cyclically applied electric fields, allows T-paths in both directions, reducing switchable polarization without causing biased loops. Such features of fatigue can be utilized to design strategies to suppress it.

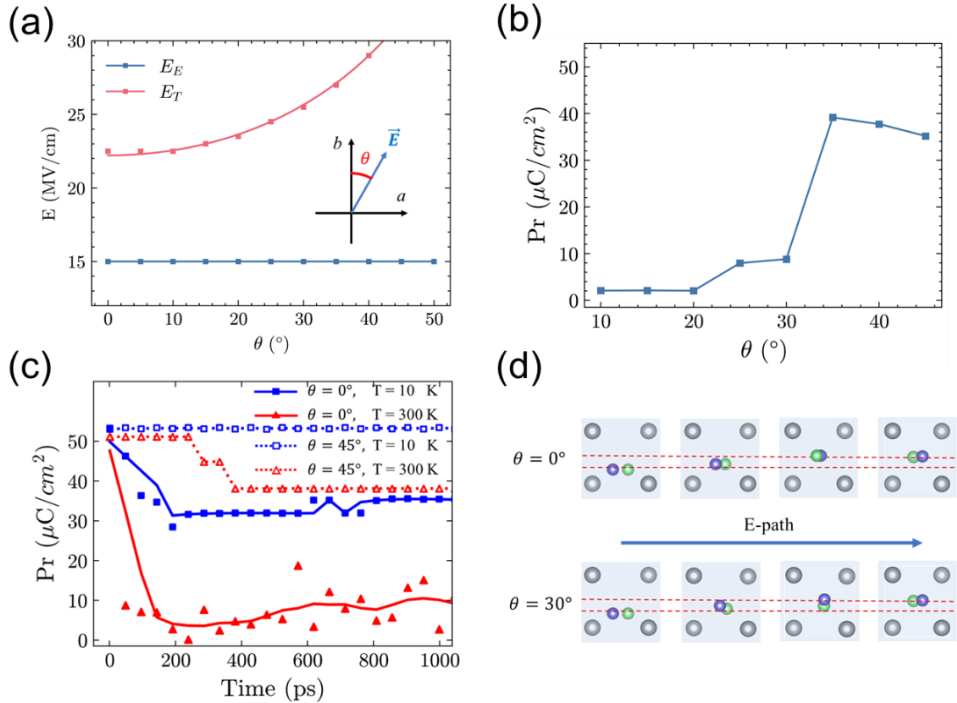

**Figure 4.** (a) $E_E$ and $E_T$ as a function of the electric field direction ($\boldsymbol{\theta}$), measured with an interval of E_step = 0.5 MV/cm. For $\boldsymbol{\theta} > 50°$, T-paths do not occur, and the polarization shifts to the *a*-direction at high electric fields instead. (b) The residual switchable polarization of $HfO_2$ after 500 ps at T = 300 K. The simulations were performed with the same supercell as in Fig. 3. (c) Residual polarization as a function of simulation time at different temperatures and values of $\boldsymbol{\theta}$ angle. (d) Schematics illustrating the oxygen atoms displacement at $\boldsymbol{\theta = 0°}$ and $\boldsymbol{\theta = 30°}$. Grey balls for hafnium, blue and green balls for different oxygens.

**Suppressing fatigue with an inclined field.** To suppress fatigue, altering the direction of the applied electric field is found to be an effective strategy. As E-path-induced T-paths, rather than pure T-paths, are primarily responsible for fatigue, there is an opportunity to disrupt this inducing process. Simulations are conducted on the same systems where fatigue is previously observed to test the effect of applying inclined fields. In these simulations, the electric field is modified to align within the *ab*-plane, defined as $\vec{E} = E_0 \cos\theta\, \vec{e}_b + E_0 \sin\theta\, \vec{e}_a$, and the temperature is set to 300 K. As shown in Fig. 4(a), when $\theta$ varies from 0° to 50°, the critical field to induce T-path ($E_T$) gradually increases, while that for E-path ($E_E$) remain unchanged. This result indicates that the undesired T-path becomes more difficult to induce under an inclined field. We then exam the polarization, which is measured along the *b*-direction. Figure 4(b) shows that, for $\theta < 20°$, the polarization after fatigue remains less than 5 μC/cm²; while for $\theta > 30°$, a residual polarization of 35 μC/cm² is observed at room temperature, which is much close to the initial 50 μC/cm². Moreover, Fig 4(c) suggests that an inclined field may suppress fatigue more efficiently than reducing temperature.

The suppression of fatigue is linked to the distinct atomic displacements of the E-path and T-path in $HfO_2$. In the E-path, oxygen atoms not only exhibit a displacement of approximately 0.7 Å along the *b*-direction, but also a displacement of approximately 0.6 Å in the *a*-direction [see top panel of Fig. 1(b)]. When the electric field is tilted, the projection of the electric field on the trajectories of certain oxygen atoms increases. These specific oxygen atoms are more readily excited to initiate movement by the electric field, thereby inducing surrounding oxygen atoms to form a complete E-path, as illustrated in Fig 4(b). Such properties make the critical electric field for the E-path ($E_E$) less sensitive to the direction of the applied electric fields ($\theta$), resulting in the flat blue line in Fig. 4(a). In contrast, the T-path is characterized by significant displacements of oxygens only in the *b*-direction and is thus only influenced by the *b*-direction component of the electric field. This dependency necessitates a higher critical electric field for the T-path ($E_T$) as the angle $\theta$ of the applied field increases. The interaction between the E-path and T-path under varying angles of the applied electric field leads to an increase in the difference between $E_E$ and $E_T$ as $\theta$ increases, mitigating the impact of E-path-induced T-paths and effectively reducing the fatigue process.

In summary, this study utilizes machine learning model to investigate the switching pathways and domains in $HfO_2$, explains the imprint and fatigue phenomena from a domain perspective and proposes a possible solution to fatigue. The E-path, which operates exclusively within domains, is essential for generating ferroelectricity; while the T-path, which can create or annihilate domain walls, is responsible for imprint and fatigue effects. It is found that E-path-induced T-paths significantly contribute to fatigue, whereas thermo-induced T-paths cause imprint. Moreover, electric field with inclined direction is found to efficiently suppress T-paths, thereby reducing fatigue and enhancing the performance and reliability of $HfO_2$-based ferroelectric devices.

Acknowledgements.

We acknowledge financial support from the National Natural Science Foundation of China (Grants No. 12204397 and No. 52372260), the Science Fund for Distinguished Young Scholars of Hunan Province of China (Grant No. 2021JJ10036), the National Key R&D Program of China (No. 2022YFA1402901), NSFC (No. 11825403, 11991061, 12188101, 12174060, and 12274082), the Guangdong Major Project of the Basic and Applied Basic Research (Future functional materials under extreme conditions--2021B0301030005), and Shanghai Pilot Program for Basic ResearchFudan University 21TQ1400100 (23TQ017).